# Guiding light with surface exciton-polaritons in atomically thin superlattices

Sara A. Elrafei[1], T. V. Raziman[1], Sandra de Vega[2],
F. Javier García de Abajo[2,3], Alberto G. Curto[1,4,5]

[1] Department of Applied Physics and Eindhoven Hendrik Casimir Institute,
Eindhoven University of Technology, 5600 MB Eindhoven, The Netherlands

[2] ICFO-Institut de Ciencies Fotoniques, The Barcelona Institute of Science and Technology,
08860 Castelldefels (Barcelona), Spain

[3] ICREA-Institució Catalana de Recerca i Estudis Avançats, 08010 Barcelona, Spain

[4] Photonics Research Group, Ghent University-imec, Ghent, Belgium

[5] Center for Nano- and Biophotonics, Ghent University, Ghent, Belgium

* Corresponding author: A.G.Curto@TUe.nl

**Two-dimensional materials give access to the ultimate physical limits of Photonics with appealing properties for ultracompact optical components such as waveguides and modulators. Specifically, in monolayer semiconductors, a strong excitonic resonance leads to a sharp oscillation in permittivity; at energies close to an exciton, the real part of the permittivity can reach high positive values or even become negative. This extreme optical response enables surface exciton-polaritons to guide visible light bound to an atomically thin layer. However, such ultrathin waveguides support a transverse electric (TE) mode with low confinement and a transverse magnetic (TM) mode with short propagation. Here, we propose that realistic semiconductor-insulator-semiconductor superlattices consisting of monolayer $WS_2$ and hexagonal boron nitride (hBN) can improve the properties of both TE and TM modes. Compared to a single monolayer, a heterostructure with a 1-nm hBN spacer improves the confinement of the TE mode from 1.2 to around 0.5 μm, whereas the out-of-plane extension of the TM mode increases from 25 to 50 nm. We propose two simple additivity rules for mode confinement valid in the ultrathin film approximation for heterostructures with increasing spacer thickness. Stacking additional $WS_2$ monolayers into superlattices further enhances the waveguiding properties. Our results underscore the potential of monolayer superlattices as a platform for visible nanophotonics with promising optical, electrical, and magnetic tunability.**

Surface polaritons are electromagnetic surface waves that allow enhanced light-matter interaction at the nanoscale. They offer new opportunities for devices such as modulators, sensors, sources, and photodetectors. These waves propagate along the interface between two materials and decay in the perpendicular (out-of-plane) direction. Such polaritons can be sustained by different types of quasiparticles in matter, like plasmons, phonons, and excitons.[1] Noble metals are common materials for supporting surface plasmon-polaritons and guiding light below the diffraction limit.[2,3] However, active tunability remains elusive because it is difficult to substantially alter the high density of free electrons in a metal.

Compared to plasmon-polaritons, surface exciton-polaritons (SEPs) are excitons that concomitantly oscillate with photons, producing a propagating surface wave bound to the interface. Exciton-polaritons have been experimentally observed in different organic[4,5] and inorganic crystals[6,7] with large absorption coefficients. For example, molecular J-aggregates of organic dyes[8,9] can sustain SEPs at room temperature and create opportunities to realize novel sensors.[10] However, SEPs in those materials still have tunability limitations. As an alternative, atomically thin materials possess extreme optical properties that can be modulated while potentially giving access to the spin and valley degrees of freedom.[11,12] Graphene and hexagonal boron nitride can indeed support plasmon- and phonon-polaritons.[1,13] These polaritons occur, however, at terahertz and mid-infrared frequencies.[14,15]

In the visible, semiconductor monolayers of transition metal dichalcogenides (TMDs) such as $WS_2$ are good candidates for guiding light using SEPs. TMD monolayers host excitons with a high oscillator strength, producing a dramatic permittivity oscillation around the exciton energy.[16] As a result, excitons in TMDs can strongly reflect electromagnetic radiation and act as atomically thin mirrors.[17–19] Interestingly, excitons can be tuned electrically, optically, magnetically, thermally, or mechanically,[19–23] opening a promising avenue for active nanophotonic devices. Several theoretical works have dealt with the excitation of SEPs in monolayers and their coupling to nearby emitters.[23–25] A report proposed that a monolayer could support SEPs and predicted confinement to within 2 μm of the monolayer with propagation lengths exceeding 100 μm.[26] Although the near-zero thickness of the monolayer can support waveguide modes, they are loosely confined to the TMD monolayer and require a symmetric refractive index medium. One possibility to increase confinement is patterning the monolayer into a photonic crystal, which has been demonstrated for suspended structures.[27] For unpatterned monolayers, however, the proximity of the guided mode to the light line complicates experimental detection due to the requirement for a perfectly symmetric optical environment with low scattering.[28] Furthermore, detection relies critically on achieving narrow excitonic linewidths, which can require cryogenic temperatures.[29]

Here, we address the fundamental challenge of guiding light bound to atomically thin semiconductors. We propose van der Waals superlattices based on semiconductor-insulator-semiconductor heterostructures to improve the propagation characteristics of surface exciton-polaritons (Figure 1a). We show the existence of both TE and TM guided modes and compare their dispersion relations in monolayers, heterostructures, and superlattices made of monolayer $WS_2$ and hexagonal boron nitride. Compared to negligible confinement in a monolayer, we demonstrate increased confinement of the TE mode in heterostructures. Then, we clarify the impact of the thickness of the spacer layer on the guided modes. In the ultrathin film approximation, we find that the decay constants of the TE and TM modes supported by heterostructures follow simple additivity rules for their constituent layers. Additionally, we investigate the electrostatic tuning of the modes. To guide experimental realizations under different excitation conditions, we investigate the differences between two approaches for solving the modes of the superlattices using either a complex in-plane wave vector, $\beta$, or a complex frequency, $\omega$. Our study thus produces specific directions to tailor and tune guided modes in semiconductor monolayer superlattices as a platform for nanoscale photonic and optoelectronic devices.

**Strong exciton oscillator strength and permittivity**

We use $WS_2$ monolayers due to their strong exciton oscillator strength and narrow linewidth, which are better than in other semiconductors at room temperature and result in a record absorption coefficient.

To retrieve the permittivity of a realistic, high-quality monolayer, we deposit a mechanically exfoliated WS$_2$ monolayer on polydimethylsiloxane (PDMS) on a glass substrate. Using PDMS as a substrate facilitates a narrow and strong exciton peak while preserving the quantum efficiency of the monolayer emission.[30] Transmittance spectroscopy shows a strong excitonic resonance with approximately 17% transmittance contrast and a narrow exciton linewidth, $\gamma_A = 22.7$ meV (Supplementary Section S1). We fit the measured transmission spectrum using transfer-matrix analysis and model the in-plane permittivity of monolayer WS$_2$ with 4 Lorentzian oscillators[16,31,32] as $\varepsilon(E) = \varepsilon_{background} + \sum_{i=1}^{i=4} f_i/(E_{i,exciton}^2 - E^2 - i\gamma_i E)$, where $\varepsilon_{background}$ is the dielectric constant in the absence of excitons, the index $i$ represents the excitonic resonances. The spectrum features peaks associated with the A and B exciton ground states, as well as the first excited state ($n$=2) of the A exciton. The peak of the C exciton at higher energies is also included in the fitting to reproduce the overall shape of the spectrum. $E_{i,exciton}$, $f_i$ and, $\gamma_i$ are the peak energy, oscillator strength, and linewidth of each exciton.

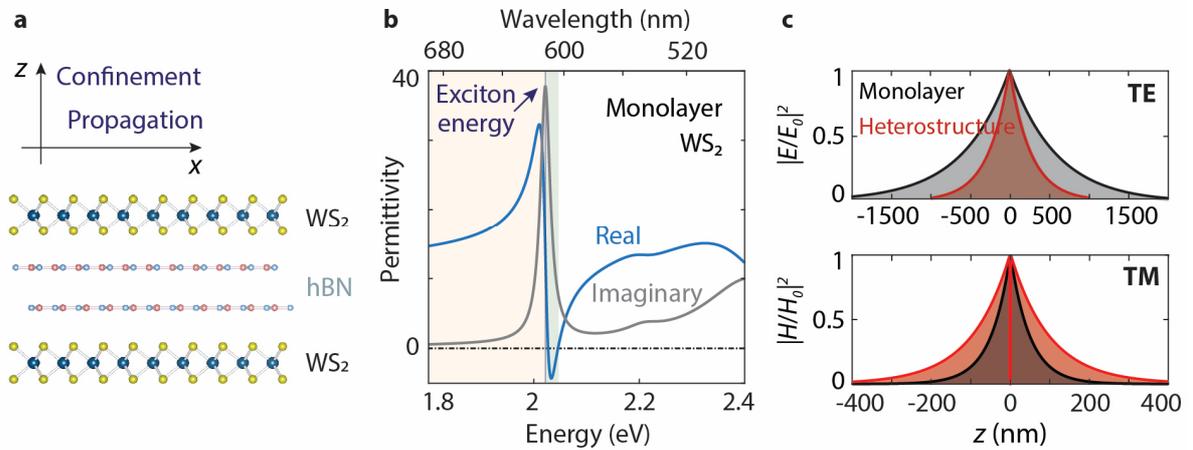

**Figure 1 | Waveguiding in WS$_2$ monolayers around a permittivity oscillation due to high exciton oscillator strength. a,** Atomically thin semiconductor-insulator-semiconductor heterostructure with a hexagonal boron nitride spacer in a symmetric refractive index environment. **b,** Experimentally retrieved permittivity of monolayer WS$_2$ obtained by fitting a transmission spectrum using the transfer-matrix method and a permittivity model with 4 Lorentzians. TE or TM modes can be supported depending on the permittivity in the yellow and green areas, respectively. **c,** Electric and magnetic field

profiles for the modes guided by a monolayer (black) and a heterostructure with a spacer thickness of 1 nm (red) at energies of 2 and 2.0223 eV. The field is confined in the out-of-plane direction, while the wave propagates in the plane. Heterostructures contribute to increased confinement of the TE mode and reduced confinement of the TM mode.

The permittivity oscillation around the exciton energy in Figure 1b is so pronounced that the real part of the permittivity, Re($\varepsilon$), goes from positive to negative across the excitonic resonance. Effectively, the material behaves optically like a high-refractive-index dielectric when Re($\varepsilon$) >0 or a reflective metal when Re($\varepsilon$) < 0. These permittivities facilitate two regimes for guiding SEP waves: a TE mode can be supported in the range of positive and high real permittivity (above 612.5 nm, orange area), while a TM mode can be sustained where the condition Re($\varepsilon(\omega)$)+ Re($\varepsilon_{medium}$) < 0 is met (from 606.5 to 612.5 nm, green area).

### Surface exciton-polaritons in monolayers and heterostructures

We consider a semiconductor monolayer as a thin film of thickness $t$ with permittivity $\varepsilon_m$ clad between two homogenous media with refractive indices $n_1$ and $n_2$. Such a layered medium can support TE and TM modes. To support a guided mode in monolayer WS$_2$, however, the environment refractive index must be nearly symmetric with $n_1 \sim n_2$. Otherwise, a cut-off appears in the minimum required TMD thickness (Supplementary Section S2). To study the mode propagation characteristics, we base our calculations on the transfer-matrix method (see Methods).[33,34] Specifically, the matrix element $M_{22}$ must be zero for a guided mode. We solve the equations numerically in the complex-$\omega$ plane to obtain the real in-plane wave vector $\beta$ of the supported guided mode (Supplementary Section S3) and evaluate its effective width $W_{eff}$ = 1/Re($q$) and effective SEP wavelength $\lambda_{SEP}$ = 2$\pi$/Re($\beta$), where $q_i = \sqrt{\beta^2 - \varepsilon_i \dfrac{\omega^2}{c^2}}$ is the momentum in the out-of-plane direction in a given medium and it is known as the decay constant. This method is appropriate for guided waves in any layered system, including atomically thin superlattices.

We use this method first to show that a $WS_2$ monolayer can support SEP modes at energies close to the exciton. SEP waves propagate along the monolayer and decay evanescently in the perpendicular direction ($z$-axis in Figure 1a). TE and TM modes can be excited in different energy ranges depending on the sign of the monolayer permittivity. The TE mode is only supported when $Re(\varepsilon_m) > 0$, while the TM mode starts to appear as the sign of the permittivity changes to negative (Figure 1b, yellow and green areas). The TE mode of a monolayer is very close to the light line (Figure 2a, black), with an effective refractive index close to the surrounding medium. Close to the exciton energy $E_A^{exc} = 2.017$ eV, the SEP wave vector becomes higher than the light line, but this mode is still only loosely confined to the monolayer. On the other hand, the TM mode is tightly confined to the monolayer, owing to the proximity of the propagation constant to the exciton energy line.

To overcome the confinement challenges associated with monolayer $WS_2$, we introduce a hexagonal boron nitride (hBN) layer with a refractive index of 2.3 between two $WS_2$ monolayers. This modification significantly alters the dispersion behavior of the TE and TM modes. In this heterostructure, the bending of the dispersion curve starts further away from the exciton energy compared to the monolayer and evolves more slowly with energy (Figure 2a, red). This enhanced mode confinement facilitates experimental observation because loosely bound guided waves are easily scattered by imperfections.

The mode profiles for a monolayer (Figure 1c, black lines) and a heterostructure (red) for the TE and TM modes at energies of 2 and 2.0223 eV, respectively, illustrate the confinement close to the monolayer. All modes show evanescent behavior outside the waveguide core, with the TM mode being more confined than the TE mode. Using a heterostructure with an hBN spacer thickness of 1 nm provides opportunities to customize the waveguide characteristics. When transitioning from a monolayer to a heterostructure at an energy of 2 eV, the effective width of the TE mode is compressed from 1.2 to approximately 0.5 µm (Figure 2b). For the TM mode at energies above the exciton peak, the mode confinement exhibits the opposite behavior and becomes less tightly confined, with the TM-mode width increasing from 47 to 90 nm for a heterostructure at an energy of 2.023 eV.

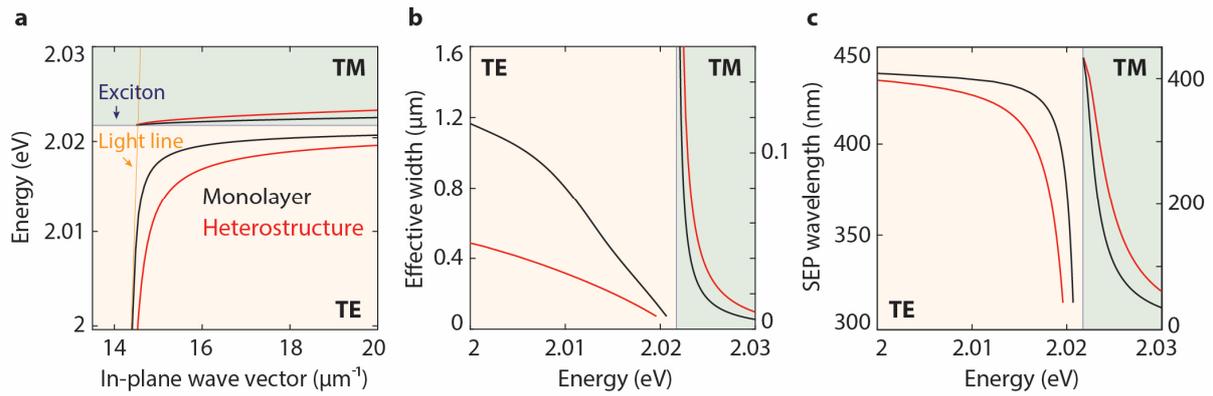

**Figure 2 | Guided modes for a WS₂ monolayer and a WS₂-hBN-WS₂ heterostructure.** Stack with spacer thickness of 1 nm in a symmetric PDMS environment. **a,** Dispersion relation for the TE (yellow area) and TM (green) modes in a monolayer (black) and a heterostructure with 1-nm-thick hBN spacer (red) calculated using the complex $\omega$ approach. Compared to the light line (orange shaded line), the TE mode is more confined for the heterostructure than for a single monolayer, while the TM mode becomes less confined. **b,** Effective width of the guided modes. **c.** Exciton-polariton wavelength as a function of photon energy.

**Contribution of the spacer to confinement**

In heterostructures, the confinement of the guided mode depends on the insulator spacer thickness. Increasing the spacer thickness typically increases the propagation constant of the TE mode. On the other hand, the TM dispersion line moves towards the monolayer curve. To gain insight into these modes, we evaluate the intensity modal profile for heterostructures with varying spacer thickness (Figure 3a). The confinement of the TE mode at $E = 2$ eV is enhanced by an order of magnitude as the spacer thickness goes from 1 to 100 nm. The TE mode shifts to higher $\beta$ as the spacer thickness increases, resulting in a more confined SEP width and a shorter SEP wavelength (Figures 3b and 3c). This apparent confinement is, however, due to the introduction of a material with a higher refractive index than the surrounding medium, which shifts the dispersion curve away from the PDMS light line towards that of the spacer material. Similarly, for the TM mode, increasing the spacer thickness reduces the width, resulting in higher confinement and shortening of the SEP wavelength (Figures 3b and 3c).

Note that the TM-mode intensity profile at $E = 2.0223$ eV (Figure 3a) corresponds to an antisymmetric electric field distribution (Supplementary Section S4).

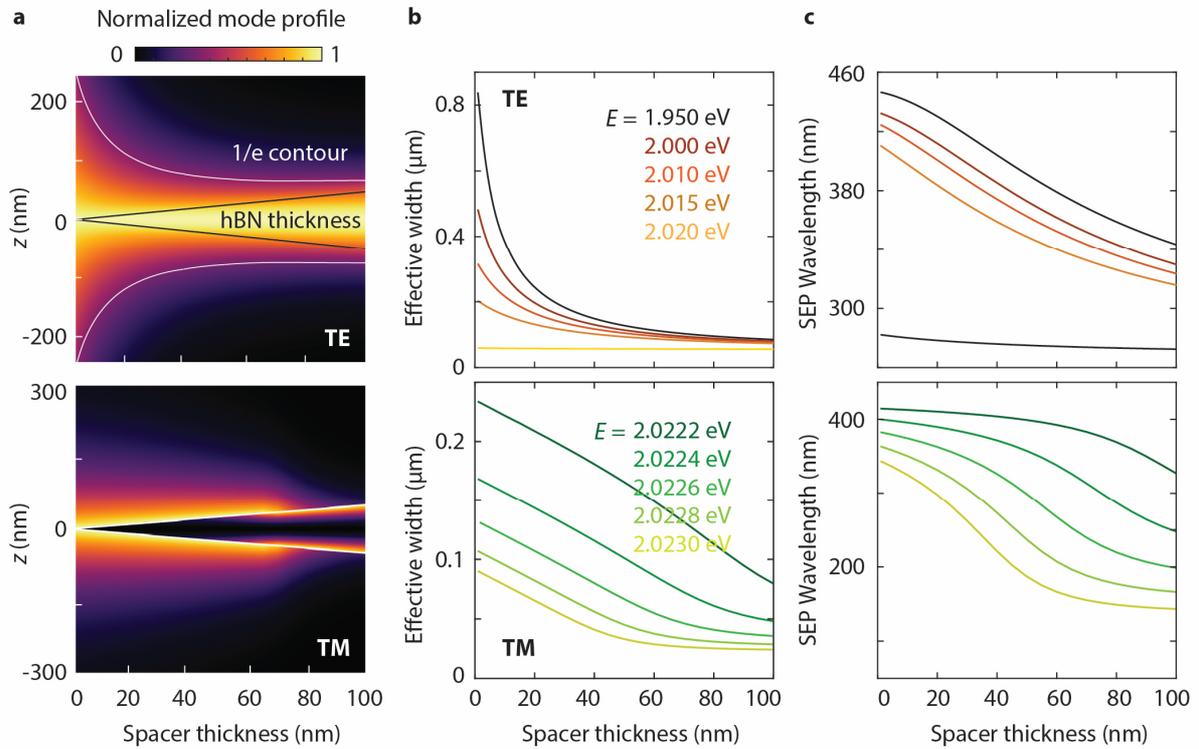

**Figure 3 | Guided mode properties in a heterostructure as a function of spacer thickness. a,** Electric field intensity profile for increasing spacer thickness for the TE mode at photon energy 2 eV and the TM mode at 2.0223 eV. The dark gray lines indicate the hBN thickness. The white contour line identifies confinement at $1/e$ intensity decay. **b,** TE- and TM-mode effective width at different energies demonstrating the contribution of the spacer thickness to confinement. **c,** Corresponding exciton-polariton wavelength, which decreases as the spacer thickens.

Next, we compare the behavior of the guided mode in a single monolayer, a heterostructure, and the hBN spacer alone, all embedded in a symmetric dielectric environment (Figure 4). Using the ultrathin film approximation, we explicitly calculate the dependence of the decay constant of the heterostructure, $q_{hetero}$, on the constituent layers for both modes. For the TE mode, the heterostructure decay constant follows a simple additivity rule of the decay constants of the individual layers, namely $q_{hBN}$ and $q_{monolayer}$, given by $q_{TE,hetero} = q_{hBN} + 2q_{monolayer}$ (proof in Supplementary Section S5). For the TM mode, on the other hand, the heterostructure decay constant is described by $q_{TM,hetero} = -2\varepsilon_{Bg}/(h\varepsilon_{hBN}$ $+ t\varepsilon_{monolayer})$, where $\varepsilon_{Bg}$ denotes the permittivity of the background medium, $h$ and $t$ represent the thicknesses of the hBN layer and the semiconductor monolayer, respectively, and $\varepsilon_{hBN}$ and $\varepsilon_{monolayer}$ are their corresponding permittivities (proof in Supplementary Section S5). These two additivity rules for TE and TM modes demonstrate the simple but distinct relations between the permittivities and thicknesses of the constituent layers and confinement in heterostructures.

We analyze first the behavior of the decay constant for the WS$_2$ monolayer and hBN layers alone. The decay constant for the monolayer is a horizontal black line in Figures 4a and 4b, as there is no spacer. If we consider an hBN film only, it supports a TE mode with increasing confinement for increasing thickness (gray line in Figure 4a). Conversely, the TM mode is absent for hBN alone at this photon energy due to its positive refractive index (no gray line in Figure 4b). For complete heterostructures containing both WS$_2$ and hBN, we observe an excellent agreement between the decay constants obtained using the analytical additivity rules (red lines in Figures 4a and 4b) and the numerically simulated decay constants (dark red), particularly for small thicknesses below a few tens of nanometers.

**Engineering the guided modes in superlattices**

A superlattice geometry − a heterostructure stack − can further improve the mode confinement and make the SEP properties more appealing for nanophotonics. The TE mode moves away from the light line for superlattices, providing higher confinement for an increasing number of monolayers (Figures 5a and 5b). The effective TE-mode width is 1.2, 0.5, and 0.3 μm for one, two, and three monolayers, respectively. The TM-mode width at $E = 2.023$ eV rises to 100 nm with three monolayers, suggesting reduced confinement of the TM mode within the structure. The dispersion line moves away from the exciton peak energy as we go from one to three monolayers (Figures 5a and 5b).

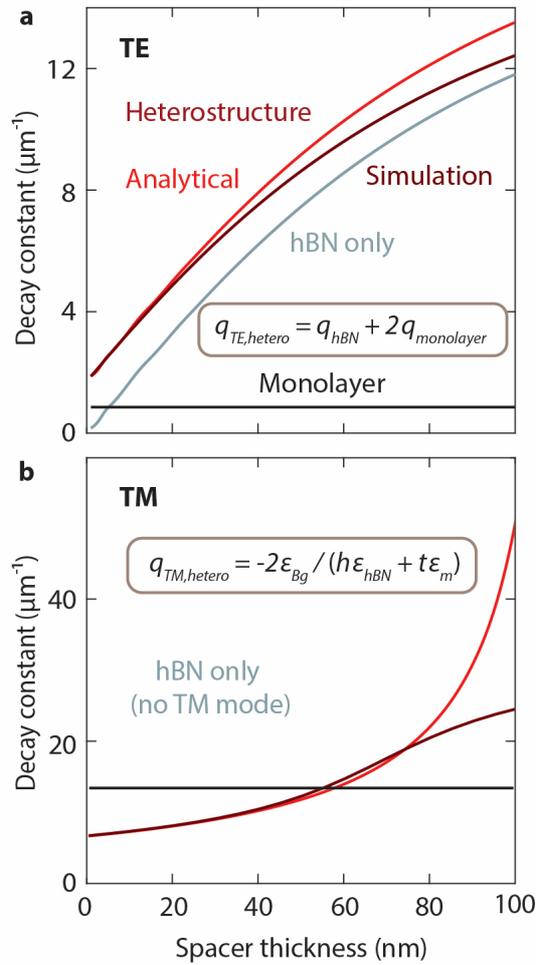

**Figure 4 | Additivity rules for the decay constants of the guided modes in a heterostructure.** Comparison between the decay constant obtained from numerical simulations (dark red) and the theoretically calculated decay constant (red) using the analytical additivity rules for **a,** TE mode at $E = 2$ eV, and **b,** TM mode at $E = 2.0223$ eV. Both show excellent agreement for thin spacers.

To exploit the advantageous tunability of SEPs, we evaluate how the electrical control of the A exciton can allow active tuning of the guided mode. The refractive index of monolayer TMDs can be tuned using electrical gating; carrier injection can tune and broaden the in-plane permittivity around the exciton resonance.[35] To incorporate tunability in our simulations, we suppress the excitonic behavior of the $WS_2$ monolayers by reducing the oscillator strength from 1.6 to 0.1 $eV^2$, resulting in a drop of ~ 50 % of its original permittivity near the A exciton (Figure 5c, inset). With this modified permittivity, we can control and potentially modulate the guided modes (Figure 5c). The TE mode confinement is frustrated in the heterostructure after suppressing the exciton by turning it towards the light line. Simultaneously, electrical tuning eliminates the possibility of sustaining the TM mode altogether because the permittivity is now positive in the energy range where a strong exciton produced a negative permittivity. Therefore, both modes show promise for modulation.

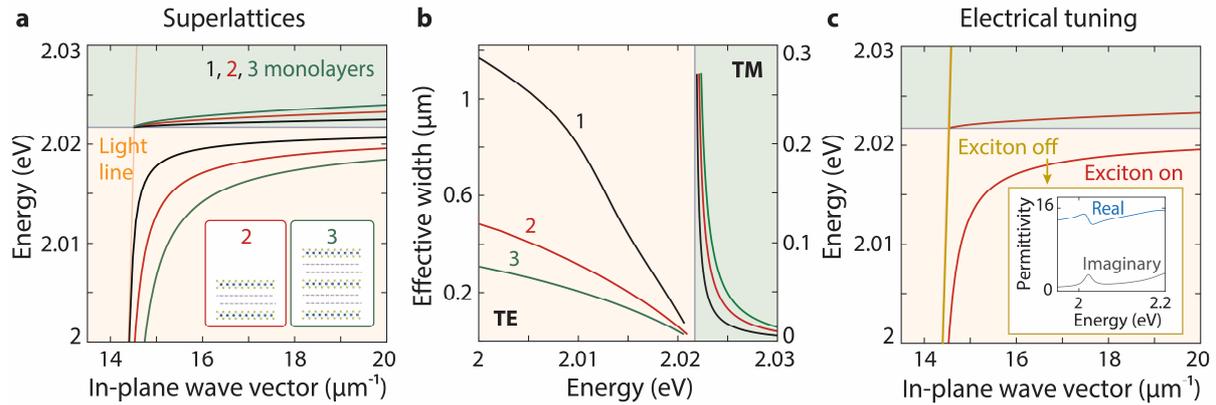

**Figure 5 | Engineering the dispersion of guided modes in superlattices. a,** Superlattices of WS$_2$ monolayers in stacked heterostructures with an hBN spacer thickness of 1 nm in a symmetric PDMS environment. In-plane wave vector as a function of energy for structures with one (black), two (red), and three monolayers (green). **b,** Comparison of their effective widths, showing a modal compression (TE) or expansion (TM) as the number of layers increases. **c,** Tunability when the exciton peak is electrically weakened in a heterostructure with two WS$_2$ monolayers with a 1-nm hBN spacer surrounded by PDMS. The TE mode evolves from a guided (red) to a nearly radiative mode (gold). The TM mode cannot be supported in the absence of a strong exciton. Inset: in-plane permittivity when the exciton is off.

**Approaches to solve the dispersion relation: complex $\beta$ and complex $\omega$**

The conditions under which SEPs can be observed in heterostructures depend on the experimental configuration. The governing equations of the guided modes are defined in the complex plane. Consequently, it is possible to solve the dispersion relation by finding the zeroes of the transfer-matrix element $M_{22}$ in the complex wave vector plane or the complex frequency plane, while keeping the other parameter real (see Methods).[36–38] The complex-$\beta$ and complex-$\omega$ approaches lead to different dispersion relations and describe different experimental conditions for polariton excitation.[37] The complex $\beta$ approach is suitable when the excitation is a monochromatic wave localized in space,[36,39–41] whereas a complex $\omega$ describes better pulsed or broadband excitation at a fixed angle.[42–46] In all our results so far, we solved the guided modes using the complex-$\omega$ approach. Here, we compare the dispersion relations obtained using the complex-$\beta$ and complex-$\omega$ approaches.

For the complex-$\omega$ solutions, we observe an asymptote for large values of $\beta$ for the TE and TM modes (Figure 6a, gray lines). Instead, for the complex-$\beta$ approach at a given real $\omega$ (purple lines), the dispersion relation of the TE mode shows a back-bending limiting the maximum value of $\beta$. We remove the unphysical branches from the complex-$\beta$ results because physically meaningful solutions should have real and imaginary parts of the wave vector with the same sign. The TM-mode dispersion lines occur within different energy ranges with a shift between complex-$\beta$ and complex-$\omega$ solutions. The reason for this shift is that when we keep $\omega$ as a real value and solve the mode equation, the obtained $\beta$ values possess a significant imaginary part for the TM mode. However, enforcing a real $\beta$ requires the film permittivity to be strongly negative. Negative permittivity only occurs near the exciton, which shifts the obtained real part of $\omega$ closer to the exciton peak. We also compare the spacer thickness dependence of the guided modes using both approaches. While the TE mode can propagate for any spacer thickness, we obtain a cut-off thickness for the TM mode for a 1-nm hBN spacer (Supplementary Section S6). Above this thickness, the effective total permittivity of the stack becomes positive, and no TM mode is supported.

Focusing on specific energies, we observe a similar evolution of the effective mode width for both approaches (Figure 6b). For complex $\beta$, the TE mode at $E = 2$ eV is confined to around 0.55 μm for the heterostructure (red) compared to 1.15 μm for the monolayer (black), whereas the TM mode at $E = 2.03$ eV expands from 30 nm for the monolayer to 80 nm for the heterostructure. Additionally, the complex-$\beta$ approach allows us to calculate an additional SEP property: the propagation length, $L_p = 1/(2\,\mathrm{Im}(\beta))$. For the monolayer, the TE and TM modes can propagate for approximately 30 μm and 2 nm at energies of 2 and 2.03 eV, respectively (Figure 6c). For the heterostructure, the propagation length of the TE mode shortens to 5 μm at the same energy and drops rapidly as the energy gets closer to the exciton due to strong absorption related to the imaginary part of the monolayer permittivity. The TM mode propagation remains extremely dampened but improves to 4 nm.

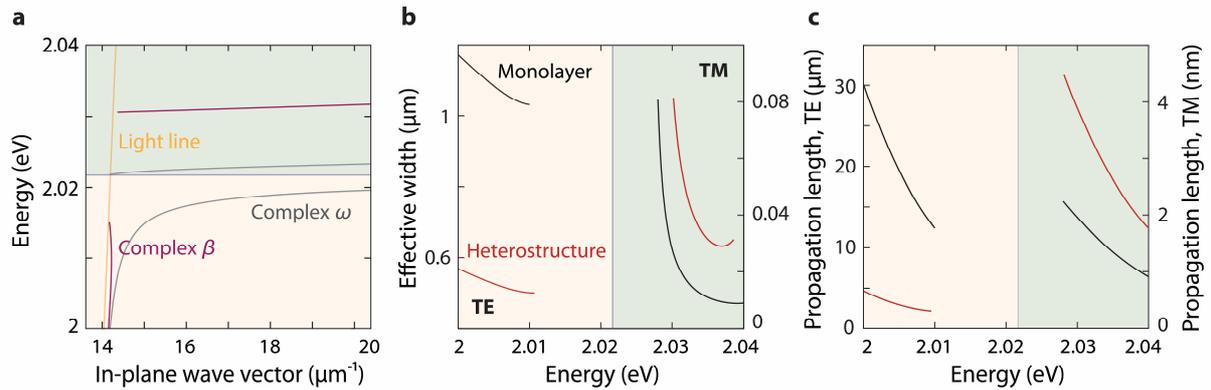

**Figure 6 | Dispersion relations in the complex-wave-vector and complex-frequency approaches. a**, Mode dispersion in a heterostructure obtained using the complex-$\beta$ approach (purple) compared to the complex-$\omega$ approach (gray), corresponding to different experimental situations. Unphysical branches are not shown in the dispersion diagram. The heterostructure consists of two WS$_2$ monolayers separated by a 0.3-nm-thick hBN monolayer in a symmetric PDMS environment. **b-c,** Effective width and propagation length using the complex-$\beta$ approach for both guided modes in the same heterostructure and in a monolayer.

Finally, we demonstrate the effect of the exciton linewidth $\gamma_A$, which can be controlled by lowering the temperature,[47,48] on the different modes in our heterostructures. We show that decreasing the linewidth (or increasing the oscillator strength) is particularly beneficial for complex-$\beta$ solutions of both TE and TM modes. We vary the A-exciton linewidth in our 4-Lorentzian permittivity model and calculate the SEP dispersion curve for $\gamma_A = 22.7$ (experimentally retrieved value at room temperature), 15, 10, and 5 meV (Figure 7). The TE mode has a more pronounced back-bending line and higher confinement using the narrowest linewidth, underscoring the need for high-quality excitons and possibly low temperatures to ease observation in experiments described by the complex-$\beta$ approach.[29] In the complex-$\omega$ approach, adjustments to the linewidth do not significantly affect the dispersion.

However, the propagation length exhibits changes because modifying the linewidth causes a shift in the permittivity in the complex plane, bringing it closer to the real axis.

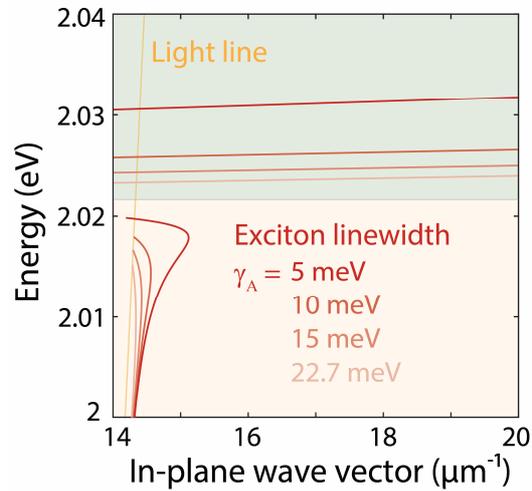

**Figure 7 | Enhanced confinement for decreasing exciton linewidth $\gamma_A$.** Exciton-polariton dispersion for a heterostructure obtained using the complex-$\beta$ approach for linewidths $\gamma_A$ = 5, 10, 15, and 22.7 meV, demonstrating the crucial role of exciton quality. The heterostructure consists of two WS$_2$ monolayers separated by a 0.3-nm-thick hBN spacer (*i.e.*, one monolayer) in a symmetric PDMS environment.

**Conclusion**

We have investigated surface exciton-polaritons supported by atomically thin semiconductor-insulator-semiconductor heterostructures and their superlattices. These guided waves rely on having strong exciton resonances with high oscillator strength and narrow linewidth, which are present in WS$_2$ monolayers even at room temperature. They also require a symmetric optical environment for observation. Both TE and TM modes are possible for high-quality monolayers within spectral ranges with positive and negative permittivities, respectively. Compared to the monolayer modes, the heterostructure architecture modifies the effective width, exciton-polariton wavelength, and propagation length. Increasing the insulator spacer thickness provides higher confinement for the TE and TM modes. Similarly, using heterostructures with more monolayers and ultrathin spacers can further increase the TE mode confinement while decreasing it for the TM mode. We proposed strongly controlling and modulating the guided modes by switching the monolayer excitons on and off. Finally, we have shown that the surface exciton-polariton waves can be predicted with either a complex wave vector or a complex frequency approach. These approaches provide qualitatively different mode dispersions and properties. As they describe different experimental conditions, it is critical to consider the right complex-plane approach to model a specific experiment.

The diverse tuning mechanisms of excitons in monolayer semiconductors provide a control knob for guided waves based on changes to the exciton strength, linewidth, and peak energy. For example, all-optical modulation due to lattice heating has been shown to substantially alter the reflectivity of atomically thin mirrors[19] and could be used to modulate exciton-polaritons in space and time. Based on our results and given the fast pace of developments in this area, atomically thin semiconductors hold great promise for nanoscale tunable Photonics at visible wavelengths.

## Methods

### Transfer-matrix method

Consider two different media separated by a planar interface. The forward and backward wave amplitudes in medium 1 are denoted by $A_1$ and $B_1$, respectively. Similarly, $A_2$ and $B_2$ are the waves in medium 2. The interface transfer matrix connects the amplitudes of the waves in the two media through

$$\begin{bmatrix} A_2 \\ B_2 \end{bmatrix} = M_{2\leftarrow 1} \begin{bmatrix} A_1 \\ B_1 \end{bmatrix}, \text{ where } M_{2\leftarrow 1} = \begin{bmatrix} M_{11} & M_{12} \\ M_{21} & M_{22} \end{bmatrix}.$$

By applying the electric and magnetic boundary conditions depending on the polarization of the wave (TE or TM), we can evaluate the matrix elements $M_{ij}$, which depend on the optical properties of the layered medium. For the TE mode, we obtain $M_{11} = \frac{1}{2k_{z2}}(k_{z2} + k_{z1})$ , $M_{12} = \frac{1}{2k_{z2}}(k_{z2} - k_{z1})$ ,

$M_{21} = \frac{1}{2k_{z2}}(k_{z2} - k_{z1})$, and $M_{22} = \frac{1}{2k_{z2}}(k_{z2} + k_{z1})$ . For the TM mode, we have $M_{11} = \frac{1}{2\frac{k_{z2}}{n_2^2}}(\frac{k_{z2}}{n_2^2} + \frac{k_{z1}}{n_1^2})$ ,

$M_{12} = \dfrac{1}{2 \dfrac{k_{z2}}{n_2^2}} (\dfrac{k_{z2}}{n_2^2} - \dfrac{k_{z1}}{n_1^2})$ , $M_{21} = \dfrac{1}{2 \dfrac{k_{z2}}{n_2^2}} (\dfrac{k_{z2}}{n_2^2} - \dfrac{k_{z1}}{n_1^2})$ , and $M_{22} = \dfrac{1}{2 \dfrac{k_{z2}}{n_2^2}} (\dfrac{k_{z2}}{n_2^2} + \dfrac{k_{z1}}{n_1^2})$ , where $k_{zi} = \pm \sqrt{\varepsilon_i \dfrac{\omega^2}{c^2} - \beta^2}$

and $\beta$ is the in-plane wave vector.

The propagation transfer matrix in a homogeneous medium is $\begin{bmatrix} A_2 \\ B_2 \end{bmatrix} = P \begin{bmatrix} A_1 \\ B_1 \end{bmatrix}$, where

$P = \begin{bmatrix} e^{ik_z d} & 0 \\ 0 & e^{-ik_z d} \end{bmatrix}$ accounts for the propagation phase, and $d$ is the thickness of the layer. The complete transfer matrix is the product of the interface and propagation matrices. For example, the transfer matrix of a film waveguide based on a monolayer is $M = M_{3 \leftarrow 2} \, P_2 \, M_{2 \leftarrow 1}$. For a heterostructure consisting of three stacked films, the transfer matrix has the form $M = M_{5 \leftarrow 4} \, P_4 \, M_{4 \leftarrow 3} P_3 \, M_{3 \leftarrow 2} \, P_2 \, M_{2 \leftarrow 1}$. We require $A_1 = B_2 = 0$ to guarantee confinement so that the field vanishes at infinity. In addition, $k_{zi}$ must have an imaginary component (and a vanishing real component if losses are neglected for the sake of determining the dispersion relation) to have decaying fields at the bottom and top layers. For a guided mode, the matrix element $M_{22}$ should be zero. Based on this condition, we can obtain the propagation constant of the mode, as well as the field distribution in each layer.

This method remains applicable for both the complex-$\omega$ and complex-$\beta$ approaches. The complex-$\omega$ approach involves finding $\omega$ for each real value of $\beta$ using a permittivity defined in the complex-$\omega$ plane. The obtained $\omega$ from the mode solution is used to extend the permittivity in the complex-$\omega$ plane using the 4-Lorentzian model (Supplementary Sections S1 and S3). Consequently, the permittivity for the complex-$\omega$ approach encompasses two branches, one for the TE mode and another for the TM mode, each requiring the determination of complex $\omega$ independently. On the other hand, the complex-$\beta$ approach relies on finding the real and imaginary parts of $\beta$ with a permittivity defined at each real $\omega$.

**Acknowledgments:** We thank Rasmus H. Godiksen, Shaojun Wang, and Ershad Mohammadi for assistance and stimulating discussions.

**Research funding:** This work was financially supported by the Netherlands Organisation for Scientific Research (NWO) through an NWO START-UP grant (740.018.009) and the Gravitation grant "Research Centre for Integrated Nanophotonics" (024.002.033).

## References

1.  Basov, D. N., Fogler, M. M. & García De Abajo, F. J. Polaritons in van der Waals materials. *Science* **354**, 6309, aag1992 (2016).

2.  Zayats, A. V., Smolyaninov, I. I. & Maradudin, A. A. Nano-optics of surface plasmon polaritons. *Physics Reports* **408**, 131–314 (2005).

3.  Weeber, J.-C. *et al.* Near-field observation of surface plasmon polariton propagation on thin metal stripes. *Phys. Rev. B* **64**, 4, 045411 (2001).

4.  Yang, F., Sambles, J. R. & Bradberry, G. W. *Long-Range Coupled Surface Exciton Polaritons*. *Phys. Rev. Lett.* **64**, 5, 559 (1990).

5.  Yang, F., Bradberry, G. W. & Sambles, J. R. Experimental Observation of Surface Exciton-polaritons on Vanadium Using Infrared Radiation. *J. Mod. Opt.* **37**, 1545–1553 (1990).

6.  Hirabayashi, I., Koda, T., Tokura, Y., Murata, J. & Kaneko, Y. Surface Exciton-Polariton in CuBr. *J. Phys. Soc. Japan* **40**, 1215–1216 (1976).

7.  Tokura, Y. & Koda, T. Surface Exciton Polariton in ZnO. *J. Phys. Soc. Japan* **51**, 2934–2946 (1982).

8.  Bradley, M. S., Tischler, J. R. & Bulović, V. Layer-by-layer J-aggregate thin films with a peak absorption constant of $10^6$ cm$^{-1}$. *Adv. Mater.* **17**, 1881–1886 (2005).

9.  Gentile, M. J., Núñez-Sánchez, S. & Barnes, W. L. Optical Field-Enhancement and Subwavelength Field-Confinement Using Excitonic Nanostructures. *Nano Lett.* **14**, 5, 2339 (2014).

10. Xu, Y., Wu, L. & Ang, L. K. Surface Exciton Polaritons: A Promising Mechanism for Refractive-Index Sensing. *Phys. Rev. Appl.* **10**, 24029 (2019).

11. Mak, K. F., He, K., Shan, J. & Heinz, T. F. Control of valley polarization in monolayer $MoS_2$ by optical helicity. *Nat. Nanotechnol.* **7**, 494–498 (2012).

12. Cao, T. *et al.* Valley-selective circular dichroism of monolayer molybdenum disulphide. *Nat. Commun.* **3**, 887 (2012).

13. Low, T. *et al.* Polaritons in layered two-dimensional materials. *Nat. Mater.* vol. 16 182–194 (2017).

14. Ju, L. *et al.* Graphene plasmonics for tunable terahertz metamaterials. *Nat. Nanotechnol.* **6**, 630–634 (2011).

15. Jablan, M., Buljan, H. & Soljačić, M. Plasmonics in graphene at infrared frequencies. *Phys. Rev. B* **80**, 245435 (2009).

16. Li, Y. *et al.* Measurement of the optical dielectric function of monolayer transition-metal dichalcogenides: $MoS_2$, $MoSe_2$, $WS_2$, and $WSe_2$. *Phys. Rev. B* **90**, 205422 (2014).

17. Back, P., Ijaz, A., Zeytinoglu, S., Kroner, M. & Imamoglu, A. Realization of an atomically thin mirror using monolayer $MoSe_2$. *Phys. Rev. Lett.* **120**, (2017).

18. Ferreira, F., Chaves, A. J., Peres, N. M. R. & Ribeiro, R. M. Excitons in hexagonal boron nitride single-layer: a new platform for polaritonics in the ultraviolet. *J. Opt. Soc. Am. B* **36**, 674 (2019).

19. Scuri, G. *et al.* Large Excitonic Reflectivity of Monolayer $MoSe_2$ Encapsulated in Hexagonal Boron Nitride. *Phys. Rev. Lett.* **120**, 037402 (2018).

20. Yu, Y. *et al.* Giant Gating Tunability of Optical Refractive Index in Transition Metal Dichalcogenide Monolayers. *Nano Lett.* **17**, 3613–3618 (2017).

21. Back, P., Zeytinoglu, S., Ijaz, A., Kroner, M. & Imamoğlu, A. Realization of an Electrically Tunable Narrow-Bandwidth Atomically Thin Mirror Using Monolayer $MoSe_2$. *Phys. Rev. Lett.* **120**, 037401 (2018).

22. Pal, A. & Huse, D. A. Many-body localization phase transition. *Phys. Rev. B* **82**, 1–7 (2010).

23. Karanikolas, V., Thanopulos, I. & Paspalakis, E. Strong interaction of quantum emitters with a

WS$_2$ layer enhanced by a gold substrate. *Opt. Lett.* **44**, 2049 (2019).

24.   Papadakis, G. T., Davoyan, A., Yeh, P. & Atwater, H. A. Mimicking surface polaritons for unpolarized light with high-permittivity materials. *Phys. Rev. Mater.* **3**, 1, 015202 (2019).

25.   Karanikolas, V. D., Marocico, C. A., Eastham, P. R. & Bradley, A. L. Near-field relaxation of a quantum emitter to two-dimensional semiconductors: Surface dissipation and exciton polaritons. *Phys. Rev. B* **94**, 19, 195418 (2016).

26.   Khurgin, J. B. Two-dimensional exciton–polariton—light guiding by transition metal dichalcogenide monolayers. *Optica* **2**, 8, 740 (2015).

27.   Zhang, X. *et al.* Guiding of visible photons at the ångström thickness limit. *Nat. Nanotechnol.* **14**, 844–850 (2019).

28.   Lee, M. *et al.* Wafer-scale δ waveguides for integrated two-dimensional photonics. *Science* **381**, 6658, 648–653 (2023).

29.   Epstein, I. *et al.* Highly confined in-plane propagating exciton-polaritons on monolayer semiconductors. *2D Mater.* **7**, 3, 035031 (2020).

30.   Eizagirre Barker, S. *et al.* Preserving the emission lifetime and efficiency of a monolayer semiconductor upon transfer. *Adv. Opt. Mater.* **7**, 13, 1900351 (2019).

31.   Wang, S. *et al.* Limits to strong coupling of excitons in multilayer WS$_2$ with collective plasmonic resonances. *ACS Photonics* **6**, 286–293 (2019).

32.   Sie, E. J. *et al.* Observation of exciton redshift-blueshift crossover in monolayer WS$_2$. *Nano Lett.* **17**, 4210–4216 (2017).

33.   Sun, B., Cai, C. & Seshasayee Venkatesh, B. Matrix method for two-dimensional waveguide mode solution. *J. Mod. Opt.* **65**, 914–919 (2018).

34.   Madrigal-Melchor, J., Pérez-Huerta, J. S., Suárez-López, J. R., Rodríguez-Vargas, I. & Ariza-Flores, D. TM plasmonic modes in a multilayer graphene-dielectric structure. *Superlattices Microstruct.* **125**, 247–255 (2019).

35.   Yu, Y. *et al.* Giant gating tunability of optical refractive index in transition metal dichalcogenide monolayers. *Nano Lett.* **17**, 3613–3618 (2017).

36.   Arakawa, E. T., Williams, M. W., Hamm, R. N. & Ritchie, R. H. Effect of damping on surface plasmon dispersion. *Phys. Rev. Lett.* **31**, 1127–1129 (1973).

37.   Archambault, A., Teperik, T. V., Marquier, F. & Greffet, J. J. Surface plasmon Fourier optics. *Phys. Rev. B* **79**, 195414 (2009).

38.   Udagedara, I. B., Rukhlenko, I. D. & Premaratne, M. Complex-ω approach versus complex-k approach in description of gain-assisted surface plasmon-polariton propagation along linear chains of metallic nanospheres. *Phys. Rev. B* **83**, 115451 (2011).

39.   Schuller, E., Falge, H. J. & Borstel, G. Dispersion curves of surface phonon-polaritons with backbending. *Phys. Lett. A* **54**, 317–318 (1975).

40.   Hu, F. *et al.* Imaging exciton-polariton transport in MoSe$_2$ waveguides. *Nat. Photonics* **11**, 356–360 (2017).

41.   Hu, F. *et al.* Imaging propagative exciton polaritons in atomically thin WSe$_2$ waveguides. *Phys. Rev. B* **100**, 121301 (2019).

42.   Wang, Q. *et al.* Direct observation of strong light-exciton coupling in thin WS$_2$ flakes. *Opt. Express* **24**, 7151 (2016).

43.   Dufferwiel, S. *et al.* Exciton-polaritons in van der Waals heterostructures embedded in tunable microcavities. *Nat. Commun.* **6**, 8579 (2015).

44.   Flatten, L. C. *et al.* Room-temperature exciton-polaritons with two-dimensional WS$_2$. *Sci. Rep.* **6**, 33134 (2016).

45.   Zhang, L., Gogna, R., Burg, W., Tutuc, E. & Deng, H. Photonic-crystal exciton-polaritons in monolayer semiconductors. *Nat. Commun.* **9**, 1–8 (2018).

46.   Kristensen, P. T., Herrmann, K., Intravaia, F. & Busch, K. Modeling electromagnetic resonators using quasinormal modes. *Adv. Opt. Photon.* **12**, 3, 612 (2020).

47.   Cadiz, F. *et al.* Excitonic linewidth approaching the homogeneous limit in $MoS_2$-based van der Waals heterostructures. *Phys. Rev. X* **7**, 2, 021026 (2017).

48.   Selig, M. *et al.* Excitonic linewidth and coherence lifetime in monolayer transition metal dichalcogenides. *Nat. Commun.* **7**, 13279 (2016).

**Supplementary Materials**

# Guiding light with surface exciton-polaritons in atomically thin superlattices

S. A. Elrafei[1], T. V. Raziman[1], S. de Vega[2],

F. J. García de Abajo[2,3], A. G. Curto[1,4,5]

[1] *Department of Applied Physics and Eindhoven Hendrik Casimir Institute,*
*Eindhoven University of Technology, 5600 MB Eindhoven, The Netherlands*

[2] *ICFO-Institut de Ciencies Fotoniques, The Barcelona Institute of Science and Technology,*
*08860 Castelldefels (Barcelona), Spain*

[3] *ICREA-Institució Catalana de Recerca i Estudis Avançats, 08010 Barcelona, Spain*

[4] *Photonics Research Group, Ghent University-imec, Ghent, Belgium*

[5] *Center for Nano- and Biophotonics, Ghent University, Ghent, Belgium*

* Corresponding author: A.G.Curto@TUe.nl

**Contents:**

**Supplementary Section S1. In-plane permittivity and transmission spectrum of WS₂**

**Sample fabrication.** We mechanically exfoliate WS₂ from a synthetic crystal (HQ Graphene) using tape (SPV 9205, Nitto Denko Co.) on optically transparent polydimethylsiloxane films (PDMS, Gel-Pak PF-80-X4) deposited on glass substrates.

**Optical measurements.** We measure the transmission spectrum at room temperature by using Köhler illumination through a microscope objective (20x Nikon CFI Plan Fluor ELWD, NA = 0.45). The signal is sent through an optical fiber to a spectrometer (Andor Shamrock 330i, with an Andor Newton 970 EMCCD camera cooled to -75 °C). We retrieve the in-plane permittivity by fitting the transmission spectrum using the transfer-matrix method and a superposition of 4 Lorentzian oscillators, namely: $\varepsilon(E) = \varepsilon_{background} + \sum_{i=1}^{i=4} f_i/(E_{i,exciton}^2 - E^2 - i\gamma_i E)$, where $\varepsilon_B$ is a background permittivity, $f_i$ is the oscillator strength of the exciton with subindex $i$, $E_{i,\,exciton}$ is corresponding the exciton peak energy, $\gamma_i$ is the linewidth of the exciton absorption band, and $E = \hbar\omega$ is the photon energy.

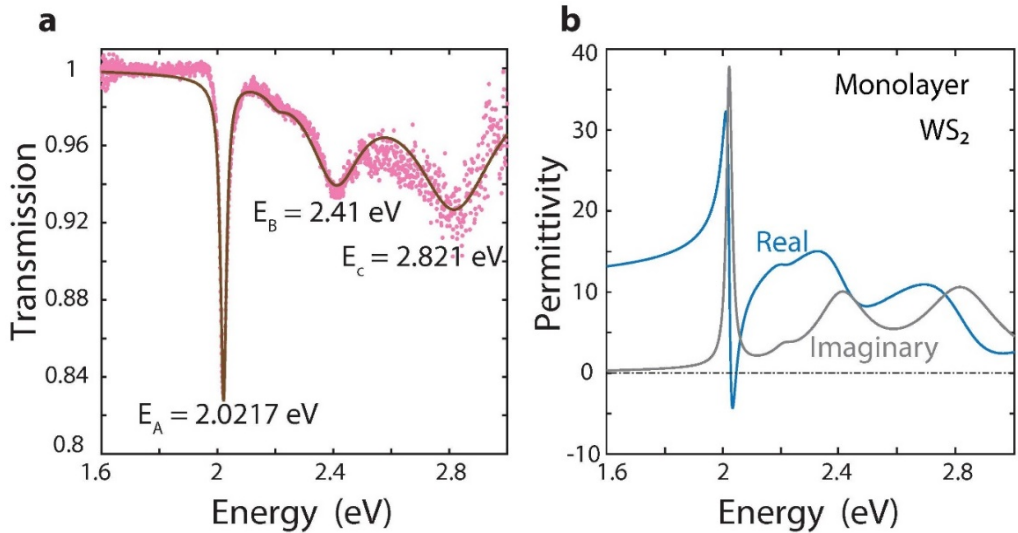

**Supplementary Figure S1 | Retrieving the in-plane permittivity of monolayer WS₂. a**, Experimental transmission spectrum of monolayer WS₂ on PDMS (pink) and fitted spectrum (brown). **b**, Retrieved in-plane permittivity of monolayer WS₂ obtained by fitting the transmission spectrum using the transfer-matrix method and a permittivity model with 4 Lorentzians.

**Supplementary Section S2. Symmetry requirement for guiding light in atomically thin waveguides**

We examine a slab waveguide model that includes a monolayer of WS$_2$, characterized by a refractive index $n_{ML}$ and surrounded by media with refractive indices $n_1$ and $n_2$ (Supplementary Figure S2a). We aim to understand how differences in these refractive indices impact the cutoff thickness for both the transverse electric (TE) and transverse magnetic (TM) modes, which is a critical parameter that defines the minimum thickness at which the waveguide can support a given mode. By solving Maxwell's equations for both modes, we find the cutoff thicknesses in Supplementary Equations S1 and S2:

$$h_{cutoff,TE}^{monolayer} = \frac{tan^{-1}[\left(\frac{n_2^2-n_1^2}{n_{ML}^2-n_2^2}\right)^{1/2}]}{(n_{ML}^2 k^2 - n_2^2 k^2)^{1/2}} \qquad \text{S1}$$

$$h_{cutoff,TM}^{monolayer} = \frac{tan^{-1}[\frac{n_{ML}^2}{n_1^2}\left(\frac{n_2^2-n_1^2}{n_{ML}^2-n_2^2}\right)^{1/2}]}{(n_{ML}^2 k^2 - n_2^2 k^2)^{1/2}} \qquad \text{S2}$$

In the case of a symmetric slab waveguide with $n_1 = n_2$, the cutoff thicknesses for the fundamental TE and TM modes are zero. We plot the relation between the cutoff thickness and the refractive index mismatch $\Delta n = n_2 - n_1$. The cutoff thickness for both TE and TM modes at the wavelengths of 615 and 608 nm decreases when $\Delta n$ is reduced, as illustrated in Supplementary Figure S2b. The cutoff thickness dependence suggests a sensitivity of waveguiding on small changes in refractive index. In the case of a single-layer-thick slab, it is crucial to limit $\Delta n$ below 5% for the TE mode to ensure light guiding. The TM mode is even more susceptible to refractive index mismatch: the mode ceases to exist for index variations of 0.006% in a PDMS environment. On the other hand, as the wavelength increases compared to the wavelengths above, the requirements for the symmetric index environment become less stringent for the TE and TM modes, thus increasing the cutoff thickness at a given refractive-index mismatch (Supplementary Figure S2c).

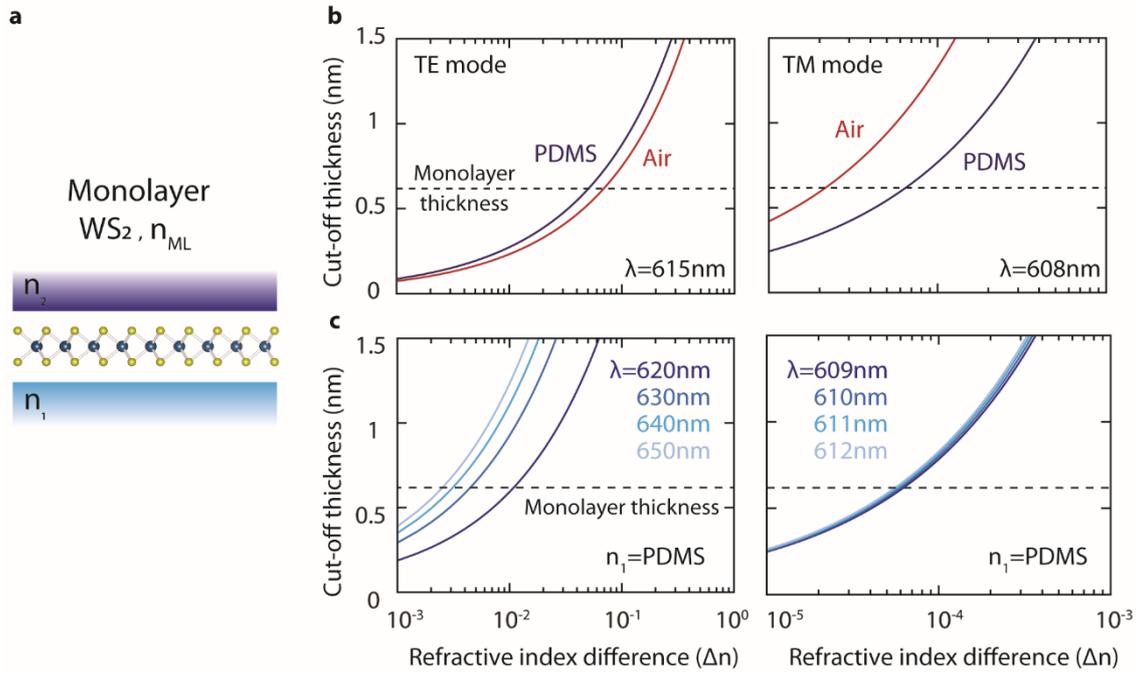

**Supplementary Figure S2 | The guided mode in a monolayer is very sensitive to asymmetries in the environment refractive index. a**, Monolayer WS$_2$ waveguide between two homogenous media with refractive indices $n_1$ and $n_2$. **b**, The cutoff thickness for guiding light depends on the difference in refractive indices ($\Delta n$) between the top and bottom cladding materials at a wavelength of 615 nm and 608 nm for TE and TM modes, respectively. Whe show results for $n_1$ corresponding to PDMS (blue) and air (red). **c**, Cutoff thickness condition at different wavelengths for both TE and TM modes when $n_1 = n_{PDMS}$.

## Supplementary Section S3. Finding the guided modes

We perform numerical simulations for a layered waveguide system using the transfer-matrix method (as described in the Methods section) for monolayers and heterostructures using different complex planes. Our dispersion calculations rely on finding the correct poles in the complex plane. To find a guided mode, we search for the roots of $M_{22} = 0$ in the complex plane, which provides the first iteration of the solution for our numerical calculation. In the complex-$\omega$ approach, we sweep the complex values of $\omega$ while keeping the wave vector $\beta$ real and fixed. The complex-$\omega$ plane shows the position of the zero at a particular value of $\beta = 15\ \mu m^{-1}$ for the TE and TM modes (Supplementary Figure S3a, top and bottom). In contrast, in the real-valued $\omega$ approach, we tune $\beta$ while $\omega$ remains constant (Supplementary Figure

S3b). Then, we use this initial solution in our simulation to evaluate the dispersion and its corresponding propagation characteristics.

This methodology remains applicable to both the complex-$\omega$ and complex-$\beta$ approaches. In the complex-$\omega$ approach, the acquired $\omega$ from the mode solution is then employed to evaluate the permittivity in the complex plane, using the 4-Lorentzian model (see above). As a result, the permittivity for the complex-$\omega$ approach encompasses two branches, one related to the TE mode and another one associated with the TM mode, each requiring an independent determination of a complex $\omega$ value (Supplementary Figure S4). On the contrary, the complex-$\beta$ approach involves the determination of the real and imaginary parts of $\beta$ with a permittivity defined at each real value of $\omega$ (Supplementary Figure S1).

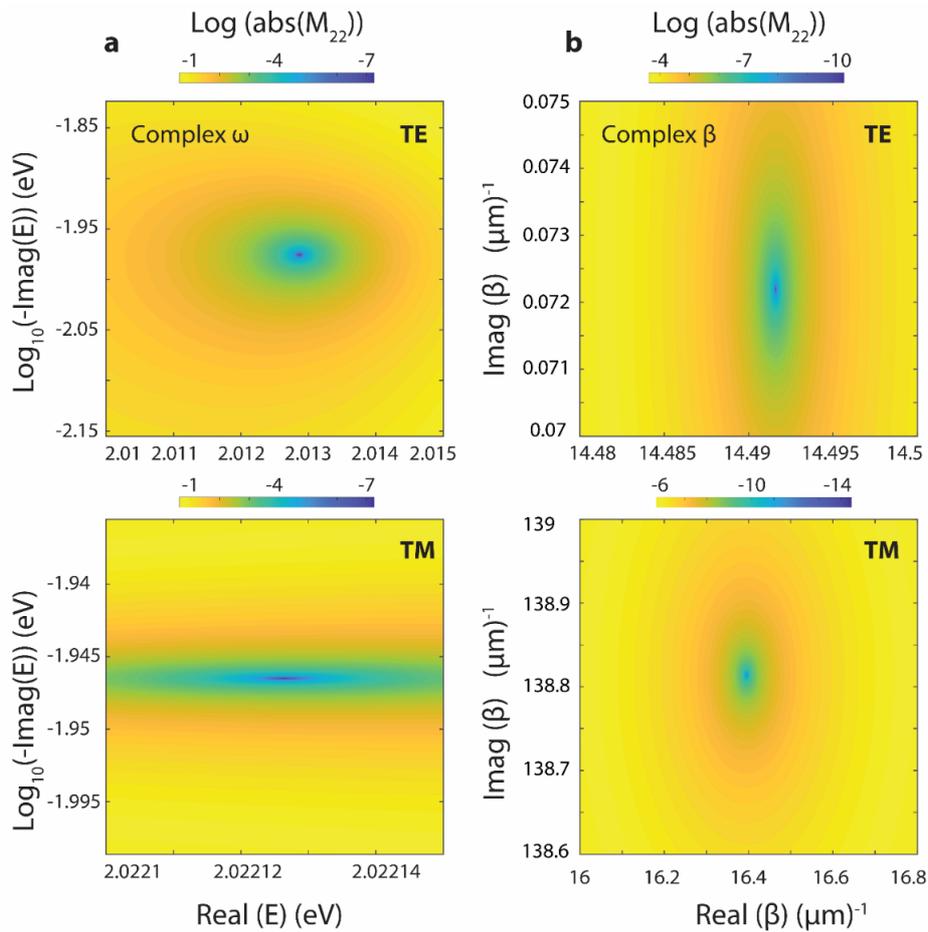

**Supplementary Figure S3 | Example of the evaluation of the first iteration solutions for TE and TM modes.** Map of the $M_{22}$ matrix element in: **a**, the complex-$\omega$ plane; **b**, the complex-$\beta$ plane.

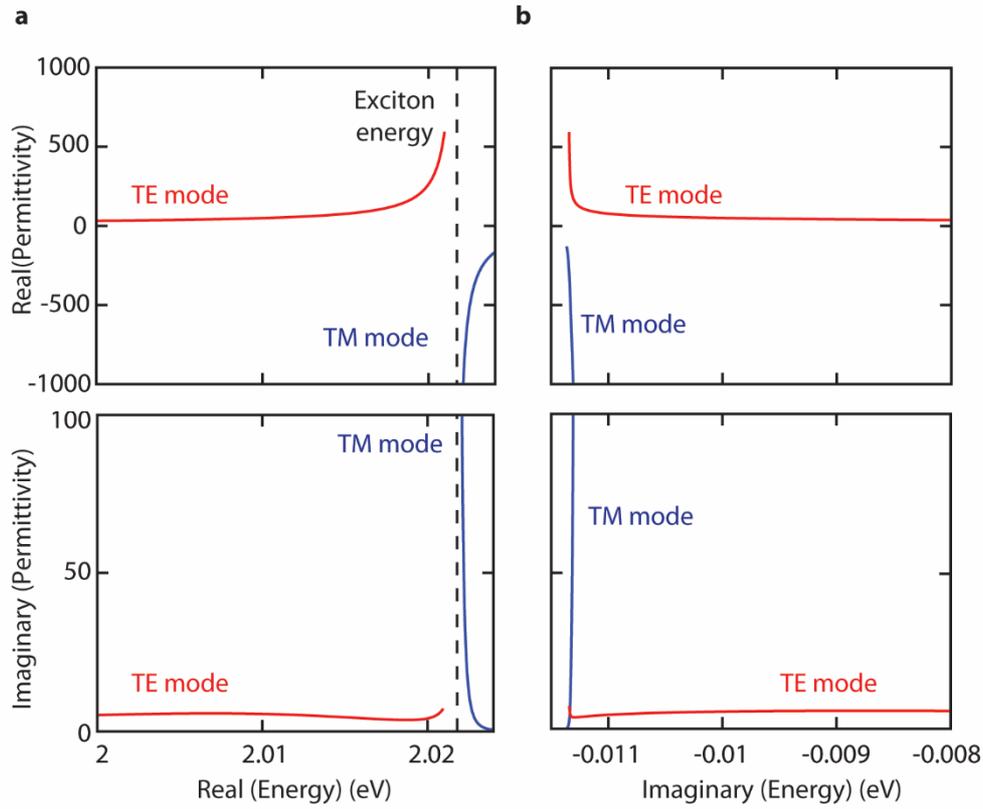

**Supplementary Figure S4 | Example of the complex permittivity of monolayer WS₂ in the complex-ω plane. a**, Real and imaginary parts of the permittivity versus the real part of the energy in the complex-ω plane for TE and TM modes. **b**, Real and imaginary parts of the permittivity versus the imaginary part of the energy for the TE and TM solutions.

## Supplementary Section S4. Field distribution for TE and TM modes in heterostructures

Using the transfer-matrix method for monolayers and heterostructures, we evaluate the field distribution of the supported TE and TM modes. These two types of waveguide modes have markedly different field distributions. For the TE mode, the normalized tangential electric field component shows a symmetric distribution at $E = 2$ eV (Supplementary Figure S5). In contrast, the TM magnetic field distribution is anti-symmetric with respect to the middle plane of the hBN spacer. This field component goes through zero at the center of the hBN film.

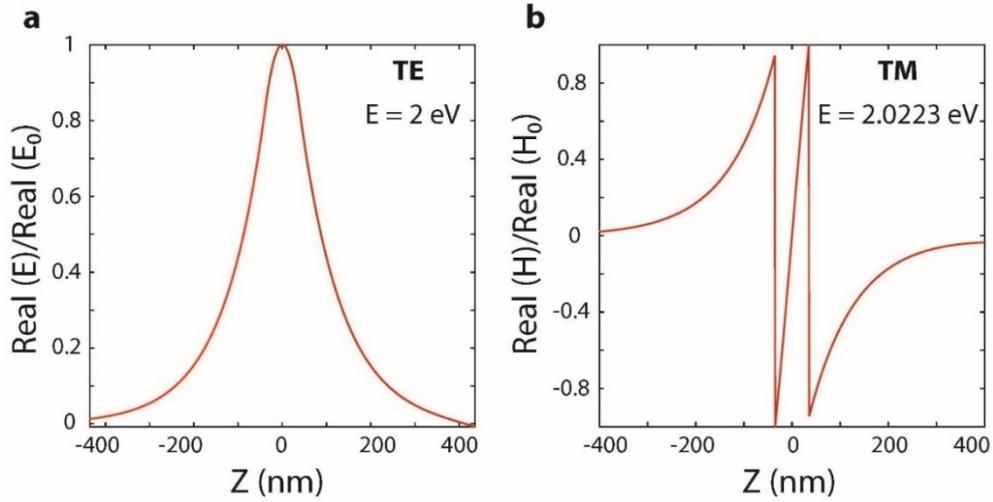

**Supplementary Figure S5 | Normalized field components for a heterostructure. a**, Normalized electric field for the TE mode at $E = 2$ eV. **b**, Normalized magnetic field for the TM mode at $E = 2.0223$ eV. The heterostructure has an hBN spacer thickness of 70 nm for ease of visualization.

## Supplementary Section S5. Additivity rules for out-of-plane momentum in heterostructures

Consider a heterostructure composed of two monolayers with thickness $t_{monolayer}$ and surface conductivity $\sigma$, separated by a spacer material with thickness $h = 2a$ and permittivity $\varepsilon_2$. This heterostructure is placed in a homogenous medium with permittivity $\varepsilon_1$ (Supplementary Figure S6).

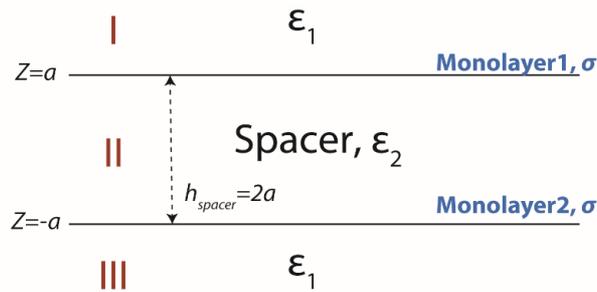

**Supplementary Figure S6 | Evaluation of decay constants.** Schematic of a heterostructure comprising two monolayers with surface conductivity $\sigma$, and a spacer material with thickness $h = 2a$ and permittivity $\varepsilon_2$. The structure is placed in a homogeneous environment with permittivity $\varepsilon_1$.

**TE mode**

For medium $n$ in the homogeneous regions in the heterostructure and using Maxwell's equations, we can write the fields:

$$\overrightarrow{E_n} = (A_n e^{i\beta x + ik_z z} + B_n e^{i\beta x - ik_z z})\hat{y},$$

$$\overrightarrow{H_n} = \frac{-i}{\omega\mu_o}\overrightarrow{\nabla} \times \overrightarrow{E_n},$$

$$\overrightarrow{H_{n//}} = \frac{i}{\omega\mu_o}\frac{\partial E_y}{\partial z} = \frac{-k_z}{\omega\mu_o}[A_n e^{i\beta x + ik_z z} - B_n e^{i\beta x - ik_z z}],$$

$$B_I = A_{III} = 0.$$

In medium *I* and *III,* as shown in Supplementary Figure S6, we have:

$$q^2 = \beta^2 - \varepsilon_1 k_0^2, \qquad\qquad k_z = iq, \qquad\qquad S3$$

where $q$ is the out-of-plane decay constant, $\beta$ is the propagation constant (or in-plane wave vector), $k_0$ is the free-space wavenumber, and $k_z$ is the component of the wave vector in the out-of-plane direction (perpendicular to the propagation direction).

In medium *II*, we have: $\qquad\qquad m^2 = \varepsilon_2 k_0^2 - \beta^2, \ \ k_z = m, \qquad\qquad S4$

where $m$ is a parameter related to the decay constants.

In the infinitesimally thin monolayer approximation, the boundary conditions at the interfaces are

$$E_+ = E_-, \qquad\qquad H_+ - H_- = \sigma\hat{n} \times E.$$

At the upper monolayer ($z = a$), we obtain:

$$A_I e^{-qa} = A_{II} e^{ima} + B_I e^{-ima},$$

$$\frac{-iq}{\omega\mu_o}A_I e^{-qa} + \frac{m}{\omega\mu_o}[A_{II} e^{ima} - B_{II} e^{-ima}] = \sigma A_I e^{-qa}.$$

At the lower monolayer ($z = - a)$, we have

$$B_{II} e^{-qa} = A_{II} e^{-ima} + B_{II} e^{ima},$$

$$\frac{-m}{\omega\mu_o}[A_{II} e^{-ima} - B_{II} e^{ima}] - \frac{iq}{\omega\mu_o}B_{III} e^{-qa} = \sigma B_{III} e^{-qa}.$$

We can then write all the boundary conditions as a matrix equation:

$$\begin{bmatrix} -1 & e^{ima} & e^{-ima} & 0 \\ -iq - \sigma\omega\mu_o & me^{ima} & -me^{-ima} & 0 \\ 0 & e^{-ima} & e^{ima} & -1 \\ 0 & -me^{-ima} & me^{ima} & -iq - \sigma\omega\mu_o \end{bmatrix} \begin{bmatrix} A_I e^{-qa} \\ A_{II} \\ B_{II} \\ B_{III} e^{-qa} \end{bmatrix} = 0.$$

For the mode to exist, the matrix determinant should be zero:

$$e^{-2ima}[-iq - \sigma\omega\mu_o - m]^2 - e^{2ima}[-iq - \sigma\omega\mu_o + m]^2 = 0,$$

$$\frac{m - iq - \sigma\omega\mu_o}{m + iq + \sigma\omega\mu_o} = \pm e^{-2ima}. \qquad\qquad S5$$

Next, we investigate the decay constant for a monolayer, an hBN film, and a heterostucture:

- **Only monolayer (no hBN):** this is equivalent to setting $a = 0$:

$$q = iq\omega\mu_o = 2q_{monolayer}. \qquad\qquad S6$$

- **Only hBN (no monolayer):** this is equivalent to setting $\sigma = 0$:

$$\frac{m - iq}{m + iq} = \pm e^{-2ima}. \qquad\qquad S7$$

The lowest-order solution, taking the positive sign, is:

$$im + q = e^{-2ima}\,(im - q),$$

$$q = im\frac{e^{-2ima} - 1}{e^{-2ikh} + 1} = m\tan(ma),$$

$$q^2 + m^2 = (\varepsilon_2 - \varepsilon_1)k_0^2 = R^2.$$

Using the small thickness approximation, Equation $S7$ is satisfied when $m \sim R$:

$$q_{hBN} = R\tan(Ra) \qquad\qquad S8$$

- **Heterostructure with two WS$_2$ monolayers and hBN as a spacer:**

Replacing Equation S6 in Equation S5, we obtain

$$\frac{m - i(q - 2q_{monolayer})}{m + i(q - 2q_{monolayer})} = \pm e^{-2ima}. \qquad\qquad S9$$

Let us define $\tilde{q} = q - 2q_{monolayer}$ and $\frac{m - i\tilde{q}}{m + i\tilde{q}} = \pm e^{-2ima}$. Solving in the same way as in Equations S7 and S8, we obtain

$$\tilde{q} = R\tan(Rh),$$

$$q = R\tan(Rh) + 2q_{monolayer} = q_{hBN} + 2q_{monolayer}.$$

**TM mode**

For medium $n$:

$$\overrightarrow{H_n} = [A_n e^{i\beta x + ik_z z} + B_n e^{i\beta x - ik_z z}]\hat{y},$$

$$\overrightarrow{E_n} = \frac{i}{\omega \varepsilon_o \varepsilon} \overrightarrow{\nabla} \times \overrightarrow{H_n},$$

$$\overrightarrow{E_n} = \frac{-i}{\omega \varepsilon_o \varepsilon} \frac{\partial H_y}{\partial z} = \frac{k_z}{\omega \varepsilon_o \varepsilon}[A_n e^{i\beta x + ik_z z} - B_n e^{i\beta x - ik_z z}].$$

We use the same approach as in the TE mode:

At the upper monolayer ($z = a$):

$$\frac{iq}{\omega \varepsilon_o \varepsilon_1} A_I e^{-qa} = \frac{m}{\omega \varepsilon_o \varepsilon_2}[A_{II} e^{ima} - B_{II} e^{-ima}],$$

$$A_I e^{-qa} - A_{II} e^{-ima} - B_{II} e^{-ima} = -\sigma \frac{iq}{\omega \varepsilon_o \varepsilon_1} A_I e^{-qa}.$$

At the bottom monolayer ($z = -a$):

$$\frac{-iq}{\omega \varepsilon_o \varepsilon_1} B_{III} e^{-qa} = \frac{m}{\omega \varepsilon_o \varepsilon_2}[A_{II} e^{-ima} - B_{II} e^{ima}],$$

$$A_{II} e^{-ima} + B_{II} e^{ima} - B_{III} e^{-qa} = \sigma \frac{iq}{\omega \varepsilon_o \varepsilon_1} B_{III} e^{-qa}.$$

Let us set $K_1 = \frac{iq}{\omega\varepsilon_0\varepsilon_1}$ and $K_2 = \frac{m}{\omega\varepsilon_0\varepsilon_2}$. We write all the boundary conditions as a matrix equation:

$$\begin{bmatrix} -1 - \frac{\sigma iq}{\omega\varepsilon_0\varepsilon_1} & e^{ima} & e^{-ima} & 0 \\ \frac{-iq}{\omega\varepsilon_0\varepsilon_1} & \frac{m}{\omega\varepsilon_0\varepsilon_2}e^{ima} & -\frac{m}{\omega\varepsilon_0\varepsilon_2}e^{-ima} & 0 \\ 0 & e^{-ima} & e^{ima} & -1 - \frac{\sigma iq}{\omega\varepsilon_0\varepsilon_1} \\ 0 & -\frac{m}{\omega\varepsilon_0\varepsilon_2}e^{-ima} & \frac{m}{\omega\varepsilon_0\varepsilon_2}e^{ima} & \frac{-iq}{\omega\varepsilon_0\varepsilon_1} \end{bmatrix} \begin{bmatrix} A_I e^{-qa} \\ A_{II} \\ B_{II} \\ B_{III} e^{-qa} \end{bmatrix} = 0.$$

For a mode to exist, the matrix determinant should be zero:

$$e^{-2ima}\left[\left(-1 - \frac{\sigma iq}{\omega\varepsilon_0\varepsilon_1}\right)\frac{m}{\omega\varepsilon_0\varepsilon_2} - \frac{iq}{\omega\varepsilon_0\varepsilon_1}\right]^2 - e^{2ima}\left[\left(-1 - \frac{\sigma iq}{\omega\varepsilon_0\varepsilon_1}\right)\frac{m}{\omega\varepsilon_0\varepsilon_2} + \frac{iq}{\omega\varepsilon_0\varepsilon_1}\right]^2 = 0,$$

$$e^{-2ima}\left[1 - \left(1 + \frac{\sigma iq}{\omega\varepsilon_0\varepsilon_1}\right)\frac{im\varepsilon_1}{q\varepsilon_2}\right]^2 - e^{2ima}\left[1 + \left(1 + \frac{\sigma iq}{\omega\varepsilon_0\varepsilon_1}\right)\frac{im\varepsilon_1}{q\varepsilon_2}\right]^2 = 0.$$

Let us also set $\frac{i\sigma q}{\omega\varepsilon_0\varepsilon_1} = \Phi$, $\frac{im\varepsilon_1}{k\varepsilon_2} = \psi$, and $(1 + \Phi)\psi = \eta$.

- **Only monolayer (no hBN)**: this is equivalent to setting $a=0$:

$$1 - (1 + \Phi)\psi = \pm(1 + (1 + \Phi)\psi).$$

Taking the positive sign solution:

$$1 - (1 + \Phi)\psi = 1 + (1 + \Phi)\psi,$$

$$(1 + \Phi)\psi = 0 \text{ with } \psi = 0 \text{ not possible,}$$

$$\Phi = -1, \; i\sigma q = -\omega\varepsilon_0\varepsilon_1, \; q = \frac{i\omega\varepsilon_0\varepsilon_1}{\sigma} = 2q_{monolayer}.$$

- **Heterostructure:**

$$e^{-2ima}[1 - \eta]^2 - e^{2ima}[1 + \eta]^2 = 0,$$

$$e^{-ima}[1 - \eta] = \pm e^{ima}[1 + \eta$$

$$\eta = \frac{e^{ima} \pm e^{ima}}{e^{ima} \pm e^{ima}}.$$

We take the positive sign and work out the $a \to 0$ limit:

$$\eta = -i\,tan(ma),$$

$$\left(1 + \frac{i\sigma q}{\omega\varepsilon_0\varepsilon_1}\right)\frac{im\varepsilon_1}{k\varepsilon_2} = -i\,tan(ma),$$

$$\frac{im\varepsilon_1}{k\varepsilon_2} + \frac{i\sigma m}{\omega\varepsilon_0\varepsilon_1} = -tan(ma),$$

$$\frac{1}{q} - \frac{1}{q_{monolayer}} = \frac{-a\varepsilon_2}{\varepsilon_1}\frac{tan(ma)}{ma} \sim \frac{-a\varepsilon_2}{\varepsilon_1},$$

$$q = \frac{1}{\frac{1}{q_{monolayer}} \frac{a\varepsilon_2}{\varepsilon_1}} = \frac{\varepsilon_1}{\frac{\sigma}{i\omega\varepsilon_0}a\varepsilon_2},$$

$$\sigma = -i\omega\varepsilon_0\varepsilon_m t_{monolayer},$$

$$q = \frac{-2\varepsilon_1}{\varepsilon_2 h + 2\varepsilon_m t_{monolayer}}.$$

To calculate the minimum thickness of the spacer, we set the denominator to zero:

$$h_{cutoff} = \frac{-2 Re(\varepsilon_m) t_{monolayer}}{\varepsilon_2}.$$

Therefore, when utilizing this analytical theory and applying the small-thickness approximation in the complex-$\beta$ plane with a real $\omega$, the cutoff length is independent of the permittivity of the substrate and superstrate, and it depends only on the properties of the spacer. We identify the following cutoff thicknesses for different spacer materials: 5.33 nm for air, 2.64 nm for PDMS, and 1 nm for hBN. Above these values, the effective total permittivity becomes positive, and the structure can no longer support the usual TM mode. The TE mode, on the other hand, has no cutoff.

## Supplementary Section S6. Influence of the insulator spacer on the guided modes for the complex-$\omega$ and complex-$\beta$ approaches

The dependence of the TE guided mode on the spacer thickness in the complex-$\beta$ approach agrees with the complex-$\omega$ one, as shown in Supplementary Figure S7a. However, the effect of increasing the spacer thickness on the TM mode in the complex-$\beta$ approach is clearly different from the complex-$\omega$ approach, as shown in Supplementary Figure S7b. To clarify this difference, we evaluate the derivative of the propagation constant with respect to the spacer thickness, $d\beta/dh$, which has the same sign as $\beta^2$ according to the previous derivation. When considering the complex-$\omega$ approach, where $\beta$ takes a real value, the quantity $d\beta/dh$ is found to be positive. This indicates that the derivative increases as the thickness of the spacer increases, thereby positively impacting the TM mode. Conversely, in the complex-$\beta$ approach, where $\beta$ is primarily an imaginary number, the derivative $d\beta/dh$ is negative. This suggests that, as the spacer thickness increases, the derivative decreases, exerting an opposite influence on the TM mode compared to solving the mode in the complex-$\omega$ plane.

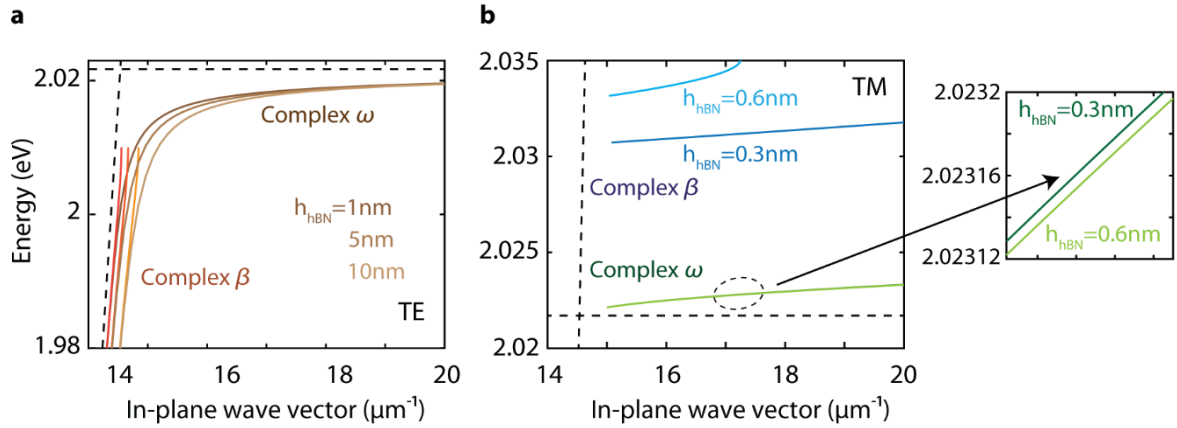

**Supplementary Figure S7 | Spacer thickness dependence of the guided modes using the complex-*ω* and complex-*β* approaches. a,** Comparison of the TE mode in the complex-*ω* and complex-*β* approaches for a heterostructure with varying hBN spacer thicknesses of 1 nm, 5 nm, and 10 nm. **b,** TM-mode dispersion in the complex-*ω* and complex-*β* approaches for spacer thicknesses of 0.3 nm and 0.6 nm.